\documentclass[prl,twocolumn,showpacs,preprintnumbers,amsmath,amssymb]{revtex4}
\usepackage{graphicx}
\usepackage{dcolumn}
\usepackage{bm}

\usepackage{amssymb}
\usepackage{epstopdf}
\newcommand{\be}{\begin{equation}}
\newcommand{\ee}{\end{equation}}
\newcommand{\bea}{\begin{eqnarray}}
\newcommand{\eea}{\end{eqnarray}}

\newcommand{\lp}{\left(}
\newcommand{\rp}{\right)}

\renewcommand{\phi}{\varphi}
\renewcommand{\epsilon}{\varepsilon}
\newcommand{\Tr}{{\rm tr}\,}
\renewcommand{\vec}[1]{{\bf #1}}

\begin{document}

\title{Flavor Symmetry and Competing Orders in Bilayer Graphene}
\author{Rahul Nandkishore and L.S. Levitov}
\affiliation{Department of Physics, Massachusetts Institute of Technology, Cambridge, MA02139}

\begin{abstract}
We analyze competition between different ordered states in bilayer graphene (BLG). Combining arguments based on SU(4) spin-valley flavor symmetry with a mean field analysis, we identify the lowest energy state with the anomalous Hall insulator (AHI). This state is an SU(4) singlet excitonic insulator with broken time reversal symmetry, exhibiting quantized Hall effect in the absence of external magnetic field. Applied electric field drives an Ising-type phase transition, restoring  time reversal symmetry. Applied magnetic field drives a transition from the AHI state to a quantum Hall ferromagnet state. We estimate energies of these states, taking full account of screening, and predict the phase diagram.
\end{abstract}

\maketitle

Bilayer graphene (BLG), due to its unique electronic structure of a two dimensional quadratically dispersing semimetal \cite{Novoselov}, offers an entirely new setting for investigating many-body phenomena. The density of states in BLG does not vanish at charge neutrality, and thus even arbitrarily weak electron interactions can trigger phase transitions in charge neutral system. Theory predicts instabilities to numerous strongly correlated gapped excitonic states with different spin and valley structure \cite{Min, Nandkishore, Zhang10}, as well as gapless nematic states \cite{Yang}.

Experiments indicate that the gapped state observed in BLG at charge neutrality in quantizing magnetic fields \cite{Feldman} persists down to low fields, crossing over to another gapped state at zero field \cite{Feldman2}. Gapped states in BLG can feature new interesting properties. Also, they can lead to new applications in nanoelectronics, in particular those based on switching between gapped and ungapped states in external fields. Hence, clarifying the relation between different gapped states and understanding the phase diagram in external fields is an interesting and timely task.

Excitonic instability results from the interaction mixing conduction and valence band states and spontaneously opening a gap between them. A variety of ordered states with different symmetries (ferromagnetic, ferrimagnetic, antiferromagnetic, ferroelectric, etc) have been analyzed using the short-range interaction model \cite{Min, Zhang10} and dynamically screened Coulomb interaction \cite{Nandkishore}. Surprisingly, all predicted gapped states appear to be degenerate on a mean field level, suggesting an underlying symmetry relation between them. Understanding the competition between these states calls for a unifying symmetry-based approach which will be outlined below.

\begin{figure}
\includegraphics[height = 1.4 in]{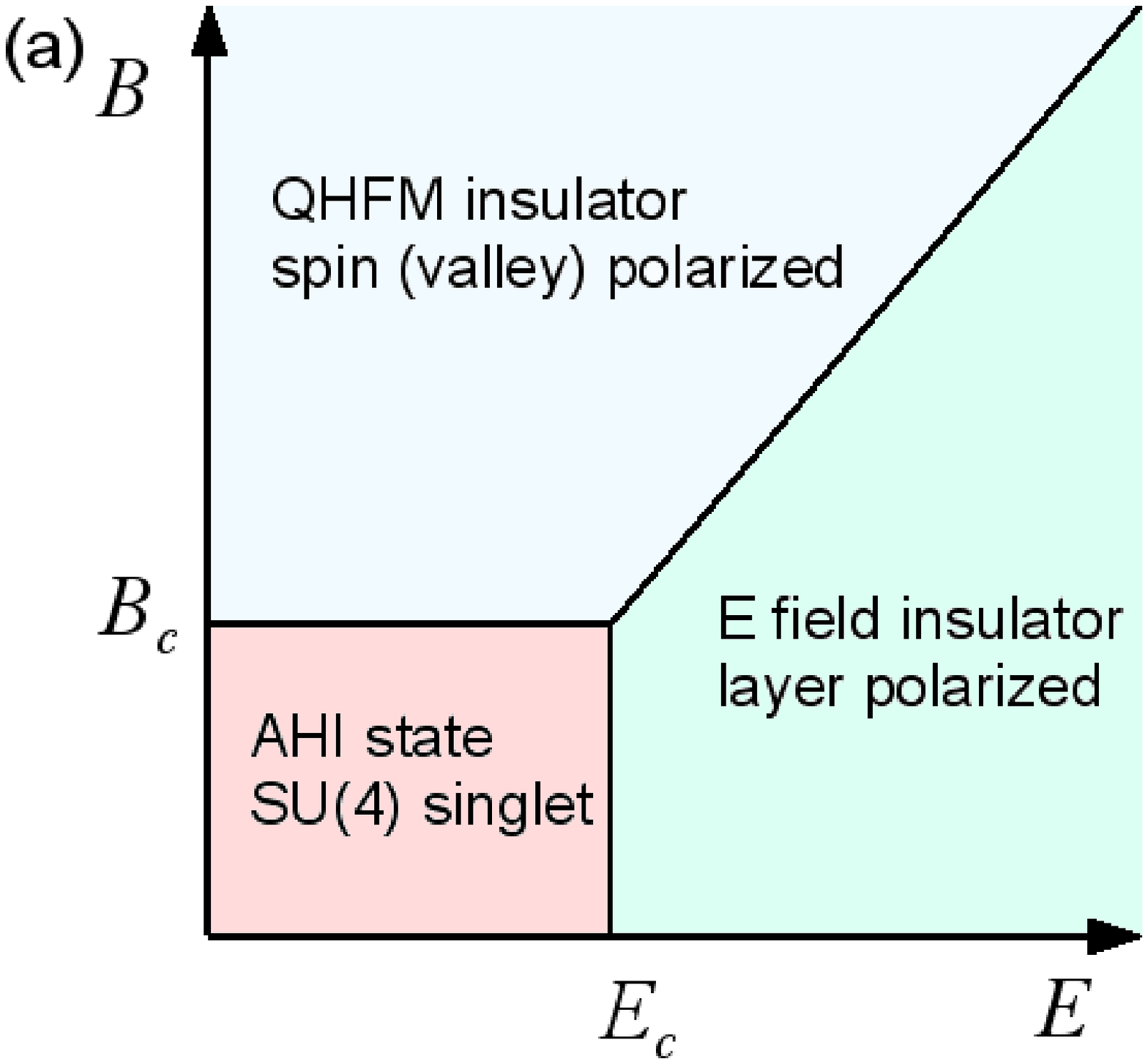}
\includegraphics[height = 1.4in]{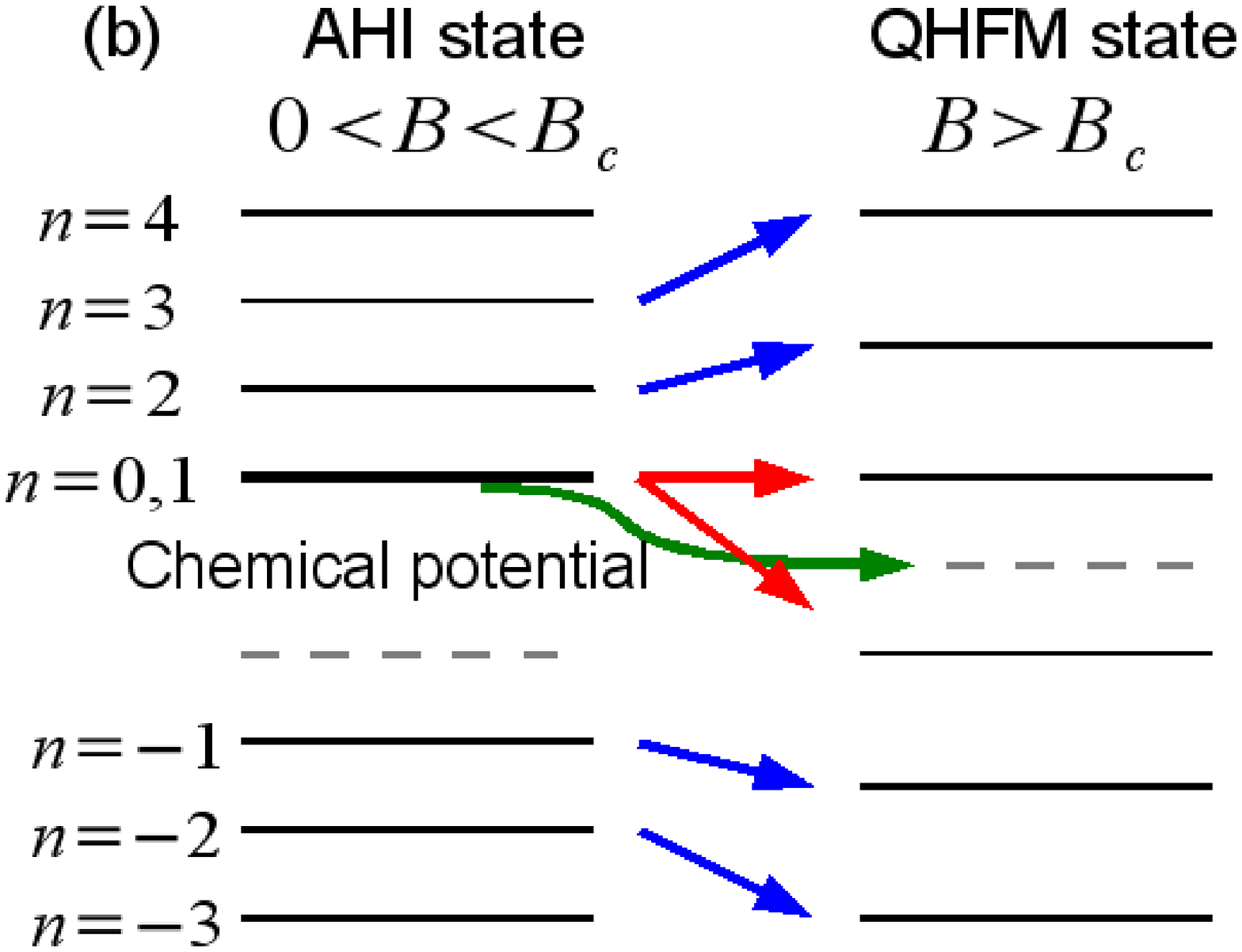}
\caption{a) Schematic phase diagram for BLG at charge neutrality. Electric and magnetic fields drive transitions from the anomalous Hall insulator (AHI) state, realized at low $E$ and $B$, to the layer polarized state (high $E$) and quantum Hall ferromagnet state (high $B$). As discussed in the text, ordering in these states is described by a $4\times 4$ matrix $Q=1$ (AHI state), $Q=\tilde \eta_3$ ($E$ field insulator), and $Q=\tilde\sigma_3$ or $Q=\tilde\eta_{1,2}$ (spin or valley polarized QHFM state). b) Landau level spectrum of the AHI and QHFM states ($E=0$). Note an anomalous Landau level in the AHI state that has no particle-hole-symmetric counterpart. Particle-hole symmetry is restored in the QHFM state, allowing the chemical potential to relax to zero.}\label{fig: phase diagram}
\vspace{-6mm}
\end{figure}

Our analysis indicates that the state realized at charge neutrality breaks $Z_2$ time reversal symmetry but is invariant under a continuous symmetry group SU(4). This state, exhibiting quantized Hall effect in the absence of magnetic field~\cite{Haldane}, is identified with one of the topological insulator states discussed recently, the anomalous Hall insulator (AHI)~\cite{Shou-Cheng}. Applied electric field drives an Ising type transition from the AHI state to a layer polarized gapped state in which time reversal symmetry is restored. Applied magnetic field drives a transition to a quantum Hall ferromagnet state [see Fig.\ref{fig: phase diagram}(a)].

As we shall see, these states are governed by SU(4) symmetry defined in terms of four spin and valley flavors. Such symmetry is peculiar to BLG: although the $2\times2$ Hamiltonians describing particles in valleys $K$ and $K'$ have different chirality, they can be mapped on each other by permuting the wavefunction components on sublattices $A$ and $B$. Importantly, since the main interactions are SU(4) invariant, all gapped states can be classified according to different subgroups of the SU(4) group which describe different patterns of spontaneous symmetry breaking. Weaker interactions can then be incorporated as external fields in a sigma model framework. As we show below, this approach can be successfully implemented to construct a full phase diagram of BLG in external electric and magnetic fields.

We start with recalling some basic facts about electronic structure of BLG. The low-energy electron states can be described by a two-component wavefunction taking values on the $A$ and $B$ sublattice of the upper and lower layer respectively \cite{Novoselov}. To analyze the structure of the Hamiltonian, it will be convenient to combine the spin and valley components in a single eight-component wavefunction  $\psi_{\alpha, s, v}(\vec{x})$, where $\alpha$ is the sublattice (layer) index. We shall use the Pauli matrices in sublattice, spin and valley space, denoted below by  $\tau_i$, $\sigma_i$ and $\eta_i$, respectively. The low energy non-interacting Hamiltonian may then be written as 
\be
\label{eq: Hamiltonian}
H_0 = \frac{(p_x + i p_y \eta_3)^2}{2m} \tau_- + \frac{(p_x - ip_y \eta_3)^2}{2m}\tau_+  +\epsilon\tau_3
\ee
where $\tau_{\pm} = \tau_1 \pm i \tau_2$. Here $m = 0.05 m_{\rm e}$ is the effective mass, and the parameter $\epsilon$ describes the effect of transverse electric field (see discussion below).

Because of the presence of $\eta_3$ in Eq.(\ref{eq: Hamiltonian}), the Hamiltonian is not invariant under rotations of valley components. To bring it to an SU(4) invariant form, we perform a unitary transformation on all operators 
\be\label{eq: transformation}
\tilde O = U O U^{\dag}
,\quad 
U = \frac{1+\eta_3}{2} + \frac{1 - \eta_3}{2} \tau_1 
\ee
This transformation mixes the layer and valley indices of the wavefunction $\psi_{\alpha, s, v}(\vec{x})$ by interchanging layers in one of the valleys. 
Defining $p_{\pm} = p_x \pm i p_y$, the transformed non-interacting Hamiltonian takes the compact form 
%
\be
\tilde H_0 = \frac{p_+^2}{2m}\tilde \tau_- + \frac{p_-^2}{2m} \tilde \tau_+ +\epsilon \tilde \tau_3 \tilde \eta_3
\label{eq: transformed Hamiltonian}
\ee
At $\epsilon=0$, this Hamiltonian is manifestly invariant under SU(4) rotations in the spin/valley flavor space.



Electron interactions can be described by a many-body Hamiltonian written in terms of $\rho_{\vec{q}} = \sum_{\vec{p}} \psi^{\dag}_{\vec{p}} \psi_{\vec{p} + \vec{q}}$ (the density summed over layers) and $\lambda_{\vec q} =  \sum_{\vec{p}} \psi^{\dag}_{\vec{p}}\tilde \tau_3 \tilde \eta_3 \psi_{\vec{p} + \vec{q}}$ (the density difference between layers). The interacting Hamiltonian, which incorporates a difference between interlayer and intralayer interaction \cite{Nilsson}, can be written as
\be\label{eq:Hint}
H =\sum_{\vec{p}} \psi^{\dag}_{\vec{p}} \tilde H_0 
 \psi_{\vec{p}} 
+\frac12 \sum_{\vec{q}} V_+ (q) \rho_{\vec{q}} \rho_{-\vec{q}} + V_- \lambda_{\vec{q}}\lambda_{-\vec{q}},
\ee
where $V_+(q) = 2\pi e^2 /\kappa q$ is the Coulomb interaction, and $V_- = \pi e^2 d/\kappa$ accounts for the layer polarization energy (here $d=3.5{\rm \AA}$ is the BLG layer separation). The $\rho\rho$ term,
which is isotropic in flavor space and thus is SU(4) invariant, dominates because $d$ is small compared to
\be\label{Bohrs_radius}
a_0 = \hbar^2\kappa/me^2 = 10 \kappa\, {\rm \AA}
,
\ee
the characteristic lengthscale set by interactions \cite{Nandkishore}. Working in the long wavelength limit, $qd\ll1$, we shall make an approximation $V_- \ll V_+$, by suppressing the $\lambda\lambda$ term which breaks the SU(4) symmetry, and restoring it at a later stage in our analysis. 



Excitonic instability, analyzed in Refs.\cite{Min, Nandkishore, Zhang10}, results in correlated states described by the BCS-like hamiltonian
\be \label{eq: full Hamiltonian}
H = \sum_{\vec{p}} \psi^{\dag}_{\vec{p}} \left(
\frac{p_+^2\tilde \tau_- +p_-^2 \tilde \tau_+}{2m}
+ \Delta \tilde \tau_3 Q \right) \psi_{\vec{p}} + \frac12 V_+ (p) \rho_{\vec{p}} \rho_{-\vec{p}}
,
\ee
with $\Delta$ the gap order parameter and $Q$ a $4\times 4$ hermitian matrix in flavor space satisfying $Q^2 = 1$. Mean field theory \cite{Min, Nandkishore, Zhang10} fixes the value of $\Delta$ up to a sign, but leaves $Q$ undetermined, since all choices of $Q$ give the same energy. 

Since hermitian matrices satisfying $Q^2 = 1$ have eigenvalues $\pm1$,
all excitonic states can be classified as ($\alpha$, $\beta$), where $\alpha$ and $\beta$ are the numbers of $+1$ and $-1$ eigenvalues of $Q$ respectively. This gives three types of states:  $(2,2)$, $(3,1)$, and $(4,0)$. There is an additional $Z_2$ symmetry associated with the overall sign of $Q$ which is absorbed into the sign of $\Delta$. 

Symmetry of these states depends on the ordering type:  ${\rm U(1)\times SU(2)}$ for $Q$ in the $(2,2)$ manifold, and SU(3) for $Q$ in the $(3,1)$ manifold. The case $(4,0)$ is special: $Q$ is a unit matrix, and the SU(4) symmetry of the Hamiltonian is not broken. This SU(4) singlet state, which polarizes layers by valley, is the valley antiferromagnet of Ref.\cite{Min}. Other states discussed in Refs.\cite{Min, Nandkishore, Zhang10} correspond to continuous symmetry breaking of type $(2,2)$ or $(3,1)$.

While within the mean field theory all choices of $Q$ yield the same energy, this degeneracy can be lifted by the effects due to order parameter fluctuations.  These fluctuations, which tend to suppress ordering, will be stronger for states  $(2,2)$ and $(3,1)$ which break continuous symmetry and thus host gapless Goldstone modes than for the SU(4) singlet state $Q=1$.
Hence we expect $\Delta_{(2,2)},\,\Delta_{(3,1)}<\Delta_{(4,0)}$. Meanwhile, the nematic state proposed in \cite{Yang} has a higher free energy even at mean field level (see supplement). We therefore predict that BLG should spontaneously order into the $Q=1$ state. 

While the $Q=1$ state does not break continuous symmetry, 
it nonetheless has interesting properties. In particular, the 
mass term $\Delta \tilde \tau_3$ in Eq.(\ref{eq: full Hamiltonian}) breaks the time reversal symmetry, implemented for the Hamiltonian, Eq.(\ref{eq:Hint}), by $T = K \tilde \eta_1 \otimes \tilde \tau_1\otimes i \tilde \sigma_2 $, where $K$ represents complex conjugation. The $Q=1$ state is thus a state of the type proposed in \cite{Haldane} as a condensed matter realization of the parity anomaly, called `anomalous Hall insulator' (AHI) in the papers on topological insulators \cite{Shou-Cheng}.
Spontaneous breaking of time reversal results in a quantized Hall effect (QHE)
\emph{even at zero magnetic field}.

The anomalous Hall effect can be understood in terms of edge states. Since the choice of sign for $\Delta$ breaks a $Z_2$ symmetry, the AHI state falls into the universality class of the Ising model. We therefore expect spontaneous formation of domains, with $\Delta(\vec{x})$ changing sign across the domain boundaries. As shown in \cite{Morpurgo}, such domain boundaries host topologically protected edge modes. In contrast to \cite{Morpurgo}, however, all edge modes have equal chirality, 
resulting in \emph{charge} QHE rather than valley QHE.

Because domains with opposite signs of $\Delta$ will have opposite $\sigma_{xy}$, the Hall conductance of a macroscopic sample will average to a value near zero. At the same time, percolating edge states will contribute to the longitudinal conductance $\sigma_{xx}$. 

We now consider applying a transverse electric field to BLG, $\epsilon\ne 0$ in Eq.(\ref{eq: transformed Hamiltonian}), which imbalances the potentials on the two layers. This reduces the symmetry group of the Hamiltonian to ${\rm SU(2)_{\rm spin} \otimes U(1)_{\rm valley}}$, with generators $\tilde \sigma_i$ and $ \tilde \eta_3$ respectively. We note that electric field favors polarizing the layers by charge, i.e. $Q = \tilde \eta_3$. This state has the same symmetries as the Hamiltonian (at $\epsilon\ne0$), and in particular preserves time reversal symmetry. Thus an electric field induces a phase transition in which time reversal symmetry is restored, and the anomalous Hall conductance disappears. This transition is analogous to restoration of $Z_2$ symmetry in a quantum Ising ferromagnet upon application of an external magnetic field \cite{Sachdev}.

We estimate the critical electric field $E_c$ by equating $\epsilon \approx \Delta_{4,0} - \Delta_{2,2}$. Here $\epsilon= \frac12 eEd/\kappa_-$, where $E$ is the electric field that would exist in the absence of graphene, and $\kappa_-$ describes intrinsic screening. The value of $\kappa_-$ can be found by following \cite{McCann2} and cutting the infrared logarithmic divergence by the gap $\Delta$. Using the value \cite{Nandkishore}
\begin{equation}
\Delta \approx 1.5 {\rm meV} \kappa^{-2}
\label{eq: Delta}
\end{equation}
%
gives $\kappa_- \approx 3$. 
For an order of magnitude estimate, we use $\Delta_{4,0}$ instead of $\Delta_{4,0} - \Delta_{2,2} $, finding $E_c  \sim 2.6 \kappa^{-1}{\rm meV /\AA}$. 

Including the interaction $V_-$, which disfavors layer polarization
by a `capacitor energy', Eq.(\ref{eq:Hint}), does not change the above conclusions. 
While penalizing the $Q = \tilde \eta_3$ state may shift the critical value $E_c$, it does not alter the qualitative picture discussed above, whereby a time reversal symmetry that is broken at $E <E_c$ is restored at $E>E_c$. Furthermore, the quantitative effect on $E_c$ should be parametrically small in $d/a_0$. 

Including trigonal warping \cite{McCann} in our Hamiltonian similarly does not alter our conclusions. The trigonal warping term is not SU(4) invariant. However, it is time reversal invariant. Thus, when trigonal warping is sufficiently weak, it does not alter ordering in the AHI state. 

The flavor symmetry framework introduced above can be applied to BLG in a magnetic field.
The main effect of magnetic field is orbital coupling, described by
$\tilde{\vec{p}} = \vec{p} - e\vec{A}$ in Eq.(\ref{eq: full Hamiltonian}), which preserves the SU(4) symmetry of the Hamiltonian. This orbital coupling causes the spectrum to split into Landau levels \cite{McCann} with an energy spacing of order $\hbar \omega_c$, where $\omega_c = eB/mc$. 
Magnetic field also couples to spin via Zeeman interaction, characterized by $E_Z=2\mu_B B$. However, since this interaction is weak, $E_Z\ll\hbar \omega_c$, we proceed by neglecting it at first.


We expect the $Q=1$ ground state introduced above to be robust to the application of a weak magnetic field. However, the $T$ non-invariance of the mass term $\Delta \tilde \tau_3$ means that the Landau level spectrum is not invariant under $B \rightarrow -B$ \cite{Jackiw}. In particular, the zeroth Landau level forms at energy $+\Delta$ only, and has no counterpart at $-\Delta$. If BLG is to remain at charge neutrality in a magnetic field, the chemical potential $\mu$ must be pulled up to the zeroth Landau level [see Fig.\ref{fig: phase diagram}(b)]. In a system with domains, meanwhile, the chemical potential will remain fixed, resulting in band bending near domain boundaries. 

At a higher magnetic field the system will transition to a quantum Hall ferromagnet (QHFM) state \cite{Barlas}, in which spontaneous flavor polarization of the zeroth Landau level is driven by exchange energy. Within our SU(4) symmetry framework, QHFM ordering
corresponds to $Q$ of type (2,2). A $Q_{2,2}$ mass term in 
Eq.(\ref{eq: full Hamiltonian}) describes gap opening in the zeroth Landau level, with spin or valley polarization
controlled by the choice of $Q$. We now review some basic properties of the QHFM state, taking full account of screening, which was neglected by \cite{Barlas}. 

When evaluating the energy gain from QHFM ordering, 
it is essential to take into account screening arising from Landau level polarization. This is so because the characteristic energy scale for interactions within the zeroth Landau level $e^2/\kappa l_B$ greatly exceeds the Landau level spacing $\hbar \omega_c$ for all experimentally relevant magnetic fields
(here $l_B = \sqrt{\hbar c / eB}$ is  the magnetic length). 
As we shall see, the QHFM energy is controlled by the behavior of the screened Coulomb interaction $\tilde V_+(q)$
in the long wavelength limit $ql_B \lesssim 1$. Applying the result of Ref.\cite{Shizuya} 
in this limit, we find
%
\begin{equation}
\label{eqn: 2d potential}
\tilde V_+(q l_B < 1) = \frac{2\pi e^2/\kappa}{q + \lambda Z q^2 l_B}
,\quad Z=N\frac{l_B}{a_0}\gg 1
,
\end{equation}
with $N=4$ and $\lambda = 0.88$  a numerical prefactor. In a wide range, $Z^{-1}\lesssim ql_B\lesssim 1$, this expression behaves as $1/q^2$, giving rise to a  \emph{logarithmic} spatial dependence  $\tilde V_+(r)$ describing field-induced `unscreening' of the screened potential discussed in Ref.\cite{DasSarma}. Such unscreening will be manifested by any state with a dipole active gap.

Additional insight into the behavior of the screened potential Eq.(\ref{eqn: 2d potential}) can be obtained by comparing to the \emph{dynamical} polarization  function found at $B=0$ and in the absence of any gap \cite{Nandkishore}. Evaluating it at frequency $\omega = \omega_c$, in the limit $ql_B<1$ we find the interaction identical to Eq.(\ref{eqn: 2d potential}), with $\lambda = 1$.


Having obtained the screened potential $\tilde V_+$, we proceed to calculate the QHFM exchange energy gain per electron $J(B)$ in a `jellium model' \cite{Laughlin} as 
\begin{equation}
\label{eqn: exchange}
J(B) = \frac{N}{2} \int  \frac{d^2p}{(2\pi)^2} \left(1-g(p)\right) \tilde V_+(p).
\end{equation}
Here, $N=4$ reflects the number of occupied states in the zeroth Landau level, and $g(p)$ is the Fourier transform of the exchange hole
$g(r) = \langle \rho_0(r) \rho_0(0)\rangle / \langle \rho_0(0)\rangle^2$, where $\rho_0$ is the electron density 
projected onto the zeroth Landau level, and the average is taken with the many body wavefunction $|\Psi\rangle$ of the system ($\langle A \rangle = \langle \Psi| A | \Psi \rangle$). 

When calculating $g(r)$, it is important to note that the zeroth Landau level is eightfold degenerate, exhibiting an accidental orbital degeneracy in addition to the fourfold flavor degeneracy. However, as was shown in \cite{Barlas}, flavor polarization is favored over orbital polarization. 
Since the QHFM exchange energy is the same for any flavor polarized state, without loss of generality 
consider a state $|\Psi\rangle$ polarized in spin.
The exchange hole $g(r)$ for such a state may be calculated by working in Landau gauge, and remembering that single particle states with different flavor are orthogonal, whereas single particle states with different orbital index or guiding center coordinate are not orthogonal. After some algebra, we obtain
\be
\label{eqn: exchange hole}
g(r) = 1 - 
e^{-\xi^2/2}\lp \xi^2-4\rp^2/32
,\quad
\xi=r/l_B
.
\ee
We note from Eq.(\ref{eqn: exchange}) and Eq.(\ref{eqn: exchange hole}) that the exchange energy comes predominantly from wavevectors $q l_B \lesssim 1$, as expected. Using the $1/q^2$ potential, we obtain, with logarithmic accuracy,
%
%
\be\label{eq: energy}
J(B) =  \frac{2\pi}{8} \frac{e^2 a_0 \ln(1+ Z)} {\kappa \lambda l_B^2} \approx 
\frac{\pi}{8\lambda}\hbar\omega_c \ln \frac{B_0}{B}
\ee
where $\Phi_0=hc/e$ is the flux quantum, and $B_0$ is given by the condition $Z = 1$. 
This is the natural result for a particle interacting with $\sim B$ other particles a distance $\sim B^{-1/2}$ apart in a logarithmic potential, and differs strongly from the $E \sim \sqrt{B}$ prediction from \cite{Barlas}. 
The difference arises because we have taken account of screening. 

The critical magnetic field $B_c$ for the field induced transition from the AHI state to the QHFM state may be estimated by equating $J(B) = \Delta_{4,0}$, which gives $B_c \sim 0.5\,{\rm T}$. This is in agreement with recent observation of a resistive state in BLG at $\nu=0$, appearing at  $B > 0.1$T \cite{Feldman}. The resistance of this state was observed to increase exponentially with magnetic field, where the exponent scaled linearly with $B/k_{\rm B}T$. This is consistent with our analysis above, assuming that transport in the QHFM state is controlled by thermal activation across the gap $J(B)$.


The effect of external electric field on the QHFM state can be conveniently analyzed in
a `sigma-model' framework, where the electric field, the `capacitance energy'  $\tilde V_-$, and the Zeeman term are treated as perturbations. 
We suppress the unimportant orbital index, and characterize the flavor structure of the QHFM state by a $Q$ matrix of type $(2,2)$.
We can write the free energy per electron 
$F = \int d^2r f[Q(\vec{r})]$, where 
\bea
 f&=& \tilde J \Tr |\nabla Q|^2+\frac{\epsilon}{4} \Tr Q \lambda_E
+v_-(\Tr Q \lambda_E)^2 + \frac{E_z}{4} \Tr Q\lambda_z
,\nonumber
\\
 && \lambda_{E} = \tilde \eta_3 \otimes 1,\quad
\lambda_z = 1 \otimes \tilde \sigma_3,\quad 
v_-=\frac{N B}{32 \Phi_0}V_-
.
 \label{eq: sigma model}
\eea
Here, $\tilde J$ is given by Eq.(\ref{eq: energy}) up to numerical prefactor of order unity, 
and we recognize $\Tr Q \lambda_E$, and $\Tr Q \lambda_z$ to be the degrees of layer and spin polarization respectively. Linear $B$ dependence of the `capacitance energy' term 
arises because the density of electrons in the zeroth Landau level is proportional to $B$. The Zeeman term is relatively small,
$E_Z/v_-\approx (m/m_{\rm e}) \kappa \ll 1$, where $m_{\rm e}$ is the free electron mass, and can thus be neglected. 
The capacitance term, meanwhile is parametrically smaller than stiffness, $v_-/\tilde J\approx d/a_0\ll 1$, which ensures validity of the sigma model for not too strong electric fields.

The sigma model, Eq.(\ref{eq: sigma model}), can be used to assess stability of the QHFM state in the presence of electric field. 
The Hamiltonian, Eq.(\ref{eq:Hint}), at $E\neq 0$ has symmetry group ${\rm SU(2)_{\rm{spin}} \otimes U(1)_{\rm{valley}}}$, broken by the QHFM state 
for any choice of $Q$ other than $Q = 1\otimes \tilde \eta_3$. In the absence of capacitance and Zeeman energies, Eq.(\ref{eq: sigma model}) predicts the state $Q = 1\otimes \tilde \eta_3$ for any nonzero $E$. 
However, for small electric fields, such a state is disfavored by capacitor energy. Reinstating it in Eq.(\ref{eq: sigma model}), we see that the symmetry preserving state is only realised for $\epsilon > \epsilon_{c2}=NV_-B/\Phi_0$.
This critical field marks a second order phase transition from QHFM state to layer polarized state [diagonal line in Fig.\ref{fig: phase diagram}(a)].
We note that 
since $\epsilon_{c2}$ is less than $\tilde J$ by a factor of about $d/a_0$, the phase transition occurs within the realm of validity of the sigma model. The linear dependence $\epsilon_{c2}\propto B$ is in agreement with experiment \cite{Feldman2}.


In summary, we have argued that at small $E$ and $B$ BLG hosts an SU(4) flavor singlet AHI state which breaks time reversal symmetry, and exhibits QHE even at zero magnetic field. We have shown that the application of an electric field restores time reversal symmetry, while application of a magnetic field induces SU(4) symmetry breaking transition to QHFM state. We have calculated the properties of the QHFM state, taking full account of screening, and have predicted an electric field induced phase transition to a time reversal invariant state. 

We thank B. Feldman, J. Martin, T. Weitz and A. Yacoby for stimulating discussions and for sharing unpublished experimental data with us. This work was supported by Office of Naval Research Grant No. N00014-09-1-0724.

\end{document}